\documentclass{article}

\usepackage{graphicx}
\usepackage{bm}

\begin{document}

\title{Photoluminescence of double quantum wells: asymmetry and excitation laser wavelength effects}

\date{}

\maketitle


\author{C.A. Bravo-Vel\'{a}zquez}
\author{L.F Lastras-Mart\'{i}nez}
\author{O. Ruiz-Cigarrillo}
\author{G. Flores-Rangel} and 
\author{L.E Tapia-Rodr\'{i}guez}

Instituto de Investigaci\'{o}n en Comunicaci\'{o}n \'{O}ptica, Universidad Aut\'{o}noma de San Luis Potos\'{i}
Alvaro Obreg\'{o}n 64, 78000 San Luis Potos\'{i}, S.L.P., M\'{e}xico\\
Email Address:c.a.bravovelazquez@gmail.com;lflm@cactus.iico.uaslp.mx\\

\author{K. Biermann} and 
\author{P.V. Santos}

Paul-Drude-Institut f\"{u}r Festk\"{o}rperelektronik, Leibniz-Institut im Forschungsverbund Berlin e.V,\\ Hausvogteiplatz 5-7, 10117, Germany\\
Email Address: santos@pdi-berlin.de


\begin{abstract}

Circularly polarized photoluminescence (PL) spectroscopy measured at 19 K on GaAs/AlGaAs symmetric and asymmetric double quantum wells (DQW) is reported. The PL is obtained by exciting the sample with a circularly polarized (left or right) laser in order to create an initial unbalanced distribution of electron spins in the conduction band and, in this way, obtain the electron spin lifetime $\tau_s$.
The effects of the excitation laser wavelength were estimated by exciting with laser wavelengths of 701.0 nm, 787.0 nm, 801.5 nm and 806.5 nm. The increase of $\tau_s$ with the excitation wavelength is attributed to the lower initial quasi-momentum $\bf{k}$ of the excited carriers, which also reduces spin-orbit relaxation processes.
$\tau_s$ was found to be higher in asymmetric DQWs: this is attributed to the wider QWs in these samples, which reduces spin relaxation due to the  Dresselhaus mechanism. 
In addition, we also detected  a smaller  contribution from the Rashba mechanism by comparing  samples with  built-in electric fields of different orientations defined by doped barrier layers.

\end{abstract}

\section{Introduction}

Spin dynamics has become a focus of interest in semiconductor physics in the last years due to the ever closer need of the development of quantum devices.~\cite{Hirohata(2020)} It is important to have an efficient control of spin currents and the manipulation and detection of spin states.~\cite{Hirohata(2020)} For the latter, the value of the spin relaxation time $\tau_s$ is very important. The spin-orbit coupling (SOC) of carriers in a semiconductor plays an important role in the spin dynamics and, in particular, in the spin relaxation time $\tau_s$. SOC induces an energy splitting of the spin states that depends on the quasi-momentum $\bf{k}$ of the carriers. In a quantum well (QW) or a double quantum well (DQW) consisting of two QWs separated by a thin tunneling barrier based on GaAs/AlGaAs structures, the main mechanisms that induce the SOC splitting are the Dresselhaus and the Rashba effects.~\cite{Dresselhaus(1955),Ganichev(2004),Luo(2009)} The relative strength of these mechanisms is quite important and allows the tuning of $\tau_s$. In our case, the thickness of the QWs in the DQW system are different, that is the DQW is asymmetric. The asymmetry and the built-in electric field in our DQW structures are, in general, very important to establish the relative strengths of the Dresselhaus and the Rashba effects.

\begin{figure}
	\includegraphics[width=1\columnwidth]{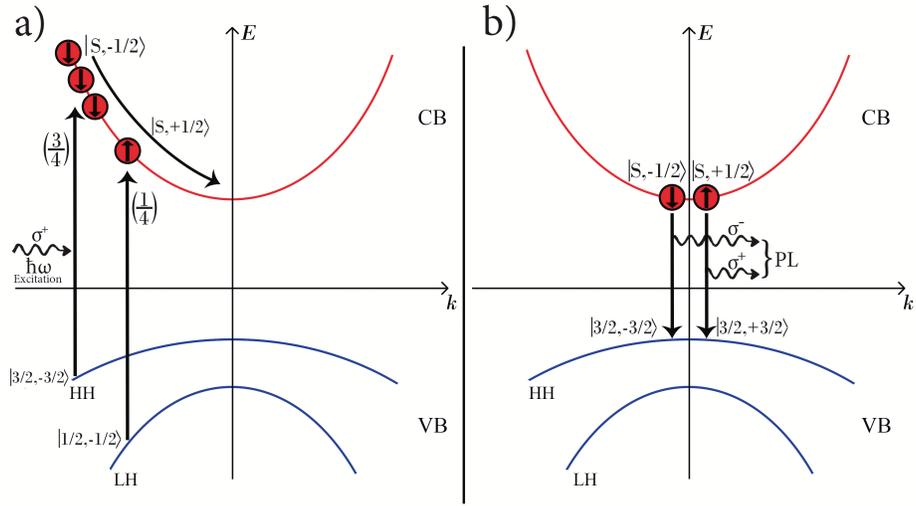}
	\caption{
	(a) Electron-hole pairs are generated by exciting with a $\sigma^+$ polarized laser. Light is absorbed with strengths of $3/4$ and $1/4$ for electron-heavy hole transitions $| 3/2~-3/2 \rangle$ and electron-light hole $| 3/2~-1/2 \rangle$ respectively. After the absorption, the spin population for $-1/2$ and $+1/2$ has a ratio of 3:1.  (b) Electrons scatter and reaches the bottom of the conduction band before recombination. The relative intensity of the PL strength for $\sigma^-$ and $\sigma^+$ polarization is associated to the population of $-1/2$ and $+1/2$ spins in the conduction band and thus can be used to study the spin dynamic during the recombination time $\tau$.}
	\label{Fig_1}
\end{figure}

Figure.~\ref{Fig_1} shows a simplified energy band diagram corresponding to a QW structure or a DQW. Due to the reduction in  symmetry from $\rm{T_d}$ to $\rm{D_{2v}}$ relative to bulk GaAs, the heavy- and light- hole levels lift their degeneracy. In the Photoluminescence (PL) setup used in this work, the sample is excited with a $\sigma^+$ polarized laser beam with an energy larger than the energy between conduction and light holes bands. Under this condition, the transitions $-3/2 \rightarrow -1/2$ and $-1/2  \rightarrow +1/2$ are excited with intensities $I\propto 3/4$ and $I\propto 1/4$ respectively.~\cite{Meier(1984a)} The difference in intensities creates an initial unbalance in the distribution of electron spins in the conduction band. If the recombination lifetime $\tau$ of the carriers is much longer than $\tau_s$ (that is, if spin relaxation occurs), the intensity of the PL signal will be the same for the $\sigma^+$ and $\sigma^-$ polarizations. In the opposite case, the difference in intensities is associated to the spin distribution of the electrons in the conduction band before recombination.

When an electron-hole pair is optically excited, the electron spin in the conduction band can be depolarized via the  Dyakonov-Perel (DP) mechanism~\cite{Dyakonov(2008),Harley(2008)} in a time $\tau_s$ before its recombination. In the DP mechanism, the electron spin precesses around an effective magnetic field caused by the Dresselhaus and the Rashba effects. The effective magnetic field and the corresponding precession frequency $\bf\Omega$ depend on the quasi-momentum $\bf{k}$ of the electrons. During $\tau$, the electron can scatter with other electrons, phonons and crystalline imperfections, and thus, in general,  $\bf\Omega$ changes randomly.

An electron in an initial spin state  $+1/2$ or $-1/2$  and an initial momentum $\bf{k}$ just after the optical excitation will suffer collisions losing its energy and reaching the bottom of the conduction band ($\bf{k}\approx 0$)  before recombination. The initial momentum  $\bf{k}$ is determined by the excitation-wavelength. During the time $\tau$, the electron follows a path in the $\bf{k}$-space, from the initial $\bf{k}$ value to the bottom of the conduction band. The effective magnetic field is proportional to $\bf{k}\cdot\bf{k}$. If the initial modulus of  $\bf{k}$ in the PL experiment is decreased, the maximum magnetic field and the depolarization of the spins will also be reduced. Thus, for excitation energies close to the band gap of the DQW, the DP mechanism can be diminished. In this way, by varying the excitation-energies, the average degree of polarization of the spins before the recombination can be tuned.

As previously mentioned, an built-in or externally applied electric field to the DQW structure induces a Rashba effect. Considering that one of the main contribution of this work is to demonstrate experimentally the dependence of the degree of circular polarization of the PL induced by the asymmetry of the DQW structure, the contribution of the built-in electric field of the structure must be quantified. To take into account this point, two types of DQW structures were studied. One with an AlGaAs n-type barrier and the other with an AlGaAs p-type barrier.   

\begin{figure}
  \includegraphics[width=1.5\columnwidth]{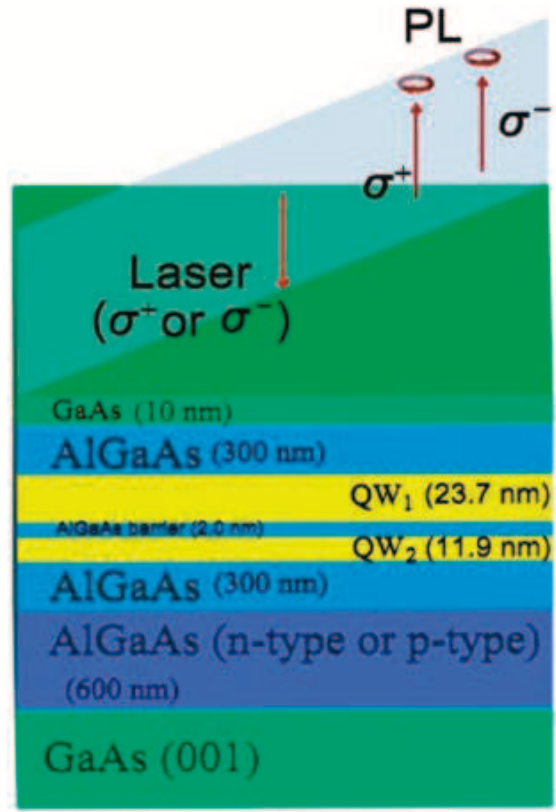}
  \caption{Structure of the asymmetric DQWs used in this work. Two samples with an AlGaAs layer of n- and p-type are studied. A circular polarized laser ($\sigma^+$ or $\sigma^-$) of different wavelengths is used to excite the sample. The PL is collected and the spectra for both $\sigma^+$ and $\sigma^-$ polarizations are measured at T=19 K.}
  \label{Fig_2}
\end{figure}

The DQW structures used in this work are shown in Fig.~\ref{Fig_2}. The $\rm{Al_{x}Ga_{1-x}As}$ layers in the structure have an Al concentration of $\rm{x=0.15}$. The QWs have thicknesses of $23.7~\rm{nm}$ and $11.9~\rm{nm}$ separated by a thin AlGaAs layer of $2.0~\rm{nm}$, constituting an asymmetric coupled system. Two structures are considered, one with the lower AlGaAs barrier composed by an AlGaAs (300 nm) and a n-type ($6\times 10^{18}\rm{cm^{-3}}$) AlGaAs (600 nm) layers and the other by an AlGaAs (300 nm) and a p-type ($5\times 10^{19}\rm{cm^{-3}}$) AlGaAs (600 nm) layers. The samples were grown by molecular beam epitaxy (MBE) on a GaAs (001) semi-insulating substrate covered with a 200-nm-thick smooth GaAs buffer layer. Since the Fermi energy at the sample surface is pinned approximately at the center of the band gap, the n- and p-doped layers will generate electric fiels of opposite directions accross the DQW structure. This fact has an impact in the strength of the PL polarization as it will be discussed. To support our results a symmetric DQW was also analyzed. In this symmetric DQW both QWs have  $11.9~\rm{nm}$ of thickness and it contains a 600 nm thick n-type ($6\times 10^{18}\rm{cm^{-3}}$) layer similar to the n-type asymmetric DQW sample described above. Table.~\ref{Table_1} summarize the main parameters of the DQWs structures.    

\section{Theory}

In two-dimensional systems, such as the asymmetric DQWs used in this work,  the spin relaxation $\tau_s$ of the electrons in the conduction band  is dominated by the Dyakonov-Perel (DP) mechanism~\cite{Dyakonov(2008),Harley(2008)}. The spin relaxation rate is given by:~\cite{Dyakonov(2008),Harley(2008)}

\begin{equation}
\label{eq1}
\frac{1}{\tau_s}={\langle\bf\Omega}\cdot{\bf\Omega\rangle}\tau_p,
\end{equation}

\noindent where ${\langle\bf\Omega}\cdot{\bf\Omega\rangle}$ is the mean of the precession vector's square magnitude in the QW plane, $\tau_s$ and $\tau_p$ are the spin and momentum relaxation times respectively. The precession vector $\bf\Omega$ is associated to the built-in effective magnetic field induced, in general, by the Dresselhaus and the Rashba effects. The Dresselhaus effect~\cite{Dresselhaus(1955),Luo(2009)} is associated to the lack of  inversion symmetry in the zincblende symmetry semiconductors (${\rm T_d}$ symmetry) while the Rashba effect is further reduced by extrinsic effects.  This occurs if an electric field is applied perpendicular to the $(001)$ surface or for asymmetric quantum wells (QWs)~\cite{Averkiev(2006),English(2013),Hao(2009),Hao(2015)}. In the later case, an asymmetry may appear in DQWs with barriers of different heights~\cite{Hao(2009),Hao(2015)}, with different chemical composition~\cite{Richards(1993)} or by triangular barriers~\cite{Averkiev(2006)} for instance. 

In principle, the relative values for the Dresselhaus and the Rashba effects (and, consequently, the value of the precession vector)  can be controlled by: i)  the built-in electric field of the DQW structure and, ii) the relative thicknesses of the QWs. The electric field induces a Rashba effect that depends linearly with the electric field and can then be reversed if the AlGaAs barrier is n- or p-type. If the DQW system is symmetric (and no  electric field is applied), only the Dresselhaus effect is important and the effective magnetic field will have no preferential in-plane orientation, leading to an efficient spin depolarization. In the case of the asymmetric DQW system, if both the  Rashba and the Dresselhaus mechanisms have similar strengths, the direction of the effective magnetic field becomes uniaxially and parallel to the DQW plane. This case leads to the formation of helical spin density waves, the called persistent spin helix (PSH)~\cite{Koralek(2009),Walser(2012),Andrei(2012)}. In this case, $\bf\Omega$ is oriented only along either the $[110]$ or the $[\bar{1}\bar{1}0]$ direction. The degree of circular polarization of the PL is defined as 

\begin{equation}
	\label{eq-1}
	{\mathcal P}=\frac{I^{\sigma^+}-I^{\sigma^-}}{I^{\sigma^+}+I^{\sigma^-}},
\end{equation}

\noindent where $I^{\sigma^+}$ and $I^{\sigma^-}$ are the PL intensity for $\sigma^+$ and $\sigma^-$ polarizations. The spin relaxation time $\tau_s$ is related to the degree of circular polarization of the PL according with~\cite{Dyakonov(2008)}:

\begin{equation}
	\label{eq0}
	{\mathcal P}=\frac{{\mathcal P}_{o}}{1+\tau/\tau_s},
\end{equation}

\noindent where $\tau$ is the recombination lifetime of the electrons and ${\mathcal P}_{o}$ is the polarization degree if no relaxation of the spin occurs. In our experiments, the polarization degree of PL is expected to have a maximum value of ${\mathcal P}_{o}=0.5$. In the limits $\tau_s << \tau$ and  $\tau_s >> \tau$, ${\mathcal P}$ tends to zero and ${\mathcal P}_{o}$ respectively.

\section{Experimental Results and discussion}

\begin{figure}
  \includegraphics[width=0.8\columnwidth]{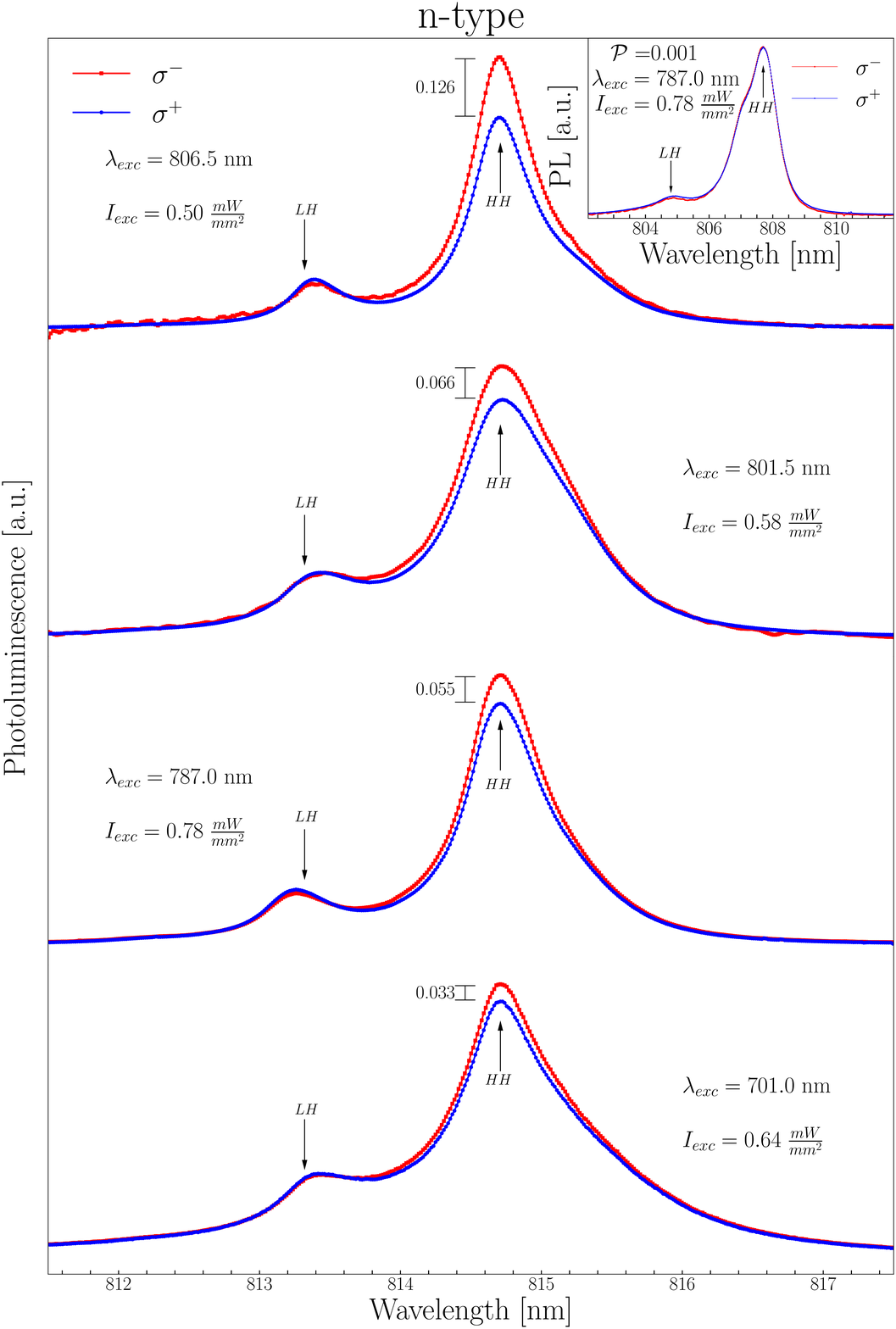}
  \caption{Photoluminescence spectra for the asymmetric DQW with a n-type AlGaAs barrier. The excitation wavelength for each spectrum is (a) 701 nm, (b) 787 nm, (c) 801.5 nm and (d) 806.5 nm. Red and blue spectra correspond to the intensity of the PL for $\sigma^+$ and $\sigma^-$ respectively. As can be seen, the degree of polarization increases with the excitation wavelength. Inset shows the spectra for the symmetric DQW excited with 787.0 nm. Note that in this case  $\sigma^+$ and $\sigma^-$ spectra  do not differ significantly.}
  \label{Fig_3}
\end{figure}

Figure.~\ref{Fig_3} shows the PL spectra for the n-type sample for excitation wavelengths of 701.0 nm, 787.0 nm, 801.5 nm and 806.5 nm. The excitation energy range lies above the electron - light-hole (LH) and the electron - heavy-hole (HH) transitions. The excitation laser has a $\sigma^+$ circular polarization state and the $\sigma^+$ and $\sigma^-$ PL curves are displayed as red and blue spectra respectively. For the HH transitions (HH in Fig.~\ref{Fig_3}), the PL intensity of the $\sigma^+$ polarization is stronger than the one for the $\sigma^-$ while for the LH transitions (LH in Fig.~\ref{Fig_3}) this relation is the opposite. This fact is expected considering that HH transitions $-1/2\rightarrow-3/2$  and $+1/2\rightarrow+3/2$ produce PL with polarizations $\sigma^-$ and $\sigma^+$, respectively, while for the LH transitions $-1/2\rightarrow+1/2$  and $+1/2\rightarrow-1/2$ produce PL with polarizations $\sigma^+$ and $\sigma^-$ respectively. Thus, the degree of circular polarization has the opposite sign for the HH and LH transitions. Note that the degree of circular polarization becomes larger (in particular, for the HH transition) when the excitation energy approaches the HH transition. This can be interpreted in the frame of the reduction of the initial quasi-momentum $\bf{k}$. With this reduction, the average magnitude of $\bf{k}$ must be smaller and, consequently, the average precession vector´s magnitude ${\langle\bf\Omega}\cdot{\bf\Omega\rangle}$, leading to an increase in the relaxation time $\tau_s$. To demonstrate that the degree of circular polarization is directly related to the asymmetry of the DQW structure, the inset of  Fig.~\ref{Fig_3} shows the PL spectra for the symmetric DQW for an excitation of 787.0 nm. In this case, 	${\mathcal P}\approx 0.001$ is more than one order of magnitude smaller than the ones obtained for the asymmetric DQW structure at the same excitation wavelength. It is important to note that for any excitation wavelength used in this work, ${\mathcal P}$ does not change significantly for the symmetric DQW.   

\begin{figure}
  \includegraphics[width=0.8\columnwidth]{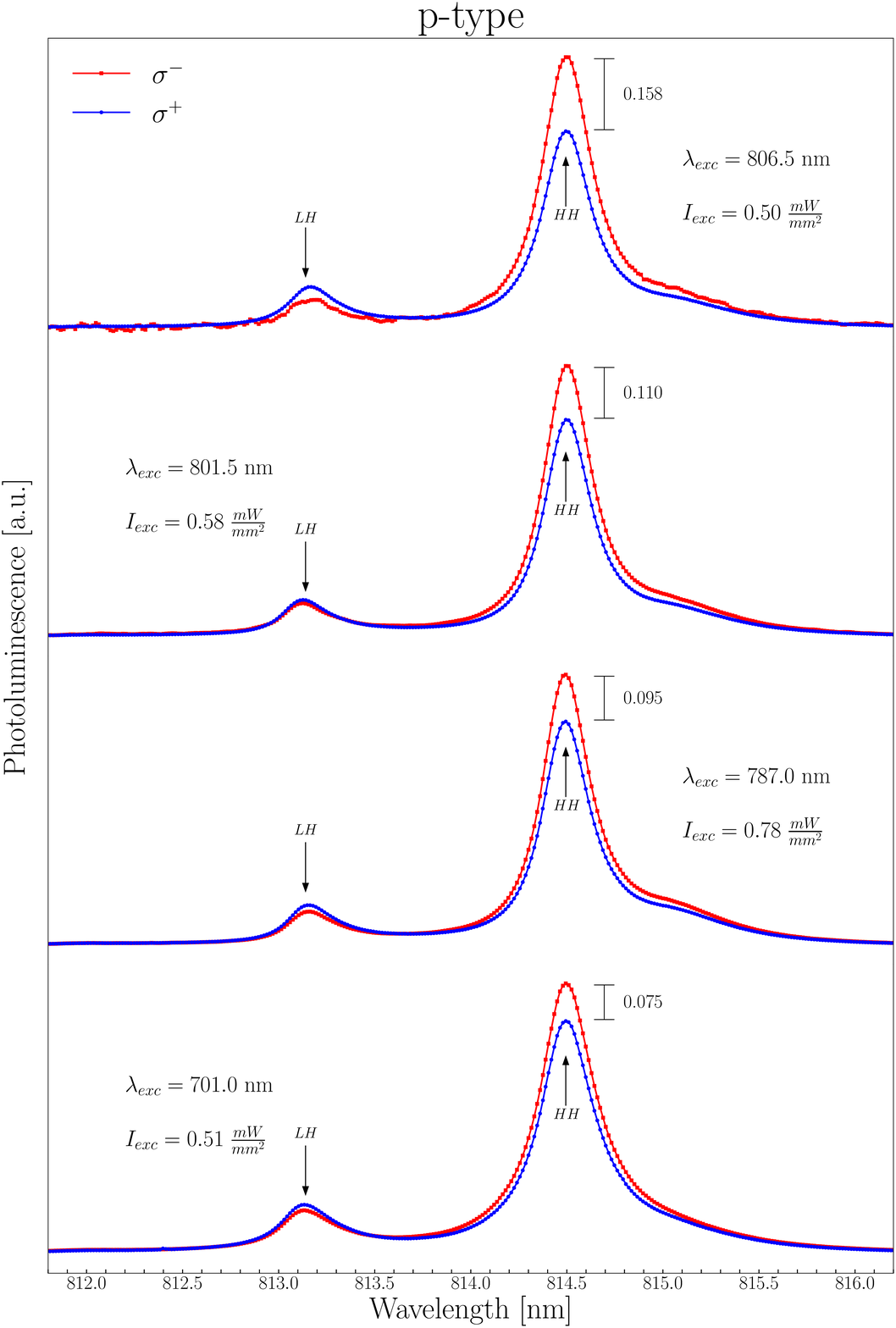}
  \caption{Photoluminescence spectra for the asymetric DQW with a p-type AlGaAs barrier. The excitation wavelength for each spectrum is (a) 701 nm, (b) 787 nm, (c) 801.5 nm and (d) 806.5 nm. Red and blue spectra correspond to the intensity of the PL for $\sigma^+$ and $\sigma^-$ respectively. As can be seen, the degree of polarization increases with the excitation wavelength.}
  \label{Fig_4}
\end{figure}

To quantify the effect of the built-in electric field on the degree of polarization of the PL, a DQW structure with p-type AlGAs barrier was measured. Figure.~\ref{Fig_4} shows the equivalent experiments of  Fig.~\ref{Fig_3}, but now performed on the p-type structure. For the PL excited with the 701.0 nm laser wavelength, the degree of circular polarization for the HH transition increases a factor of 2, while for 806.5 nm the factor it diminishes to 1.2. Considering that the electric field has opposite direction for the n- and p- type DQWs, the effect on the polarization degree is not so large. This fact suggests that the Rashba effect induced by the electric field has a small contribution in the spin depolarization. However, a Rashba effect induced by the asymmetry could have a contribution. If only the Dresselhaus effect is considered, a possible explanation to the increase of $\mathcal P$ in the asymmetric DQW, is that the PL from the asymmetric DQW arises from the wider QW, which has a smaller band gap. In QWs, the Dresselhaus effect reduces with the inverse square of thickness, increasing $1/\tau_s$. However, to clarify this point more experiments and calculations must be performed on these DQW systems.

\begin{figure}
	\includegraphics[width=1\columnwidth]{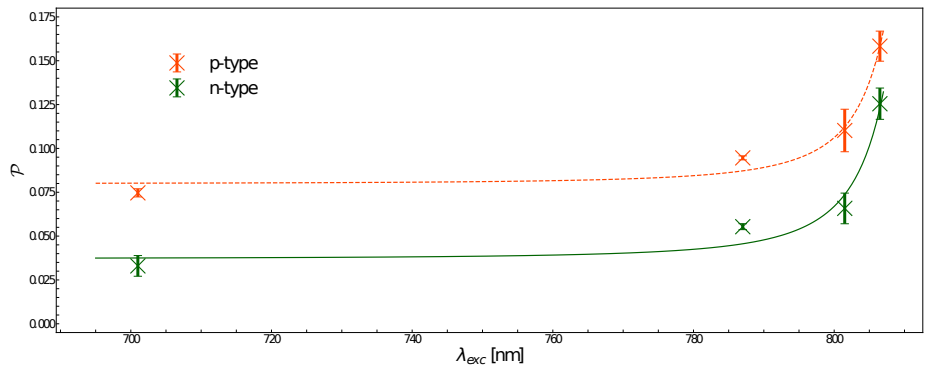}
	\caption{Polarization degree (${\mathcal P}$) of the heavy-hole transition for the asymmetric quantum well with (a) n-type and (b) p-type barriers versus the excitation wavelength. The lines are guides to the eye.}
	\label{Fig_5}
\end{figure}

The dependence of the degree of polarization $\mathcal P$ on the excitation wavelength is shown in Fig.~\ref{Fig_5}. As we mentioned at 701 nm,  $\mathcal P$ has its maximum difference for the n-type and the p-type DQW structures. When the wavelength increases, $\mathcal P$ tends to increase rapidly in both cases. This fact is consistent with the fact that when the excitation approaches to the gap of the structure (814.7 nm), the electrons in the initial $\bf{k}$ has its minimum possible value and in the same way the average precession vector´s magnitude ${\langle\bf\Omega}\cdot{\bf\Omega\rangle}$ reaches its minimum. This condition implies that electrons with lower initial  momentum have a higher probability of conserving their spin after photo-excitation. From  Eq.~\ref{eq1}, the experimental values of $\mathcal P$ and a recombination lifetime of the electrons of the order of $\tau = 1.0~ \rm{ns}$~\cite{Gobel(1983)} the relaxation time is estimated. For the n-type and p-type samples $\tau_s$ takes values in the range from 0.07 ns to 0.3 ns and  0.17 ns to 0.5 ns respectively.

\section{Conclusion}

The results reported in the present work demonstrate that the asymmetry in a DQW system increases the value of the degree of circular polarization in PL measurements. The experiments also show the dependence of the spin relaxation time of electrons in the conduction band excited by different laser wavelengths in PL measurements. By measuring the degree of circular polarization of the PL, it was demonstrated that the spin relaxation time increases for wavelengths closer in energy to the fundamental gap of the DQW system. The degree of PL circular polarization was found to be higher for asymmetric than for symmetric DQW structures. The later is attributed to the fact the the electrons are stored in a wider QW in the case  of the asymmetric samples. Furthermore, we compared  the spin lifetimes of asymmetric samples with a n- and a p- type AlGaAs lower DQW barriers and, thus, with built-in electric fields of opposite amplitudes.  $\tau_s$ was found to be slightly larger for samples with a p-doped barrier, thus demonstrating the presence of a small Rashba effect. We believe that the asymmetric DQWs constitutes an excellent system to characterize and study the spin dynamics in two-dimensional semiconductor structures.   

\section{Experimental Section}

Experiments were carried out by using a low-vibration, helium
closed-cycle cryocooler (19 K). The excitation of the samples was performed by using semiconductor lasers with wavelengths of $\lambda$=~701~nm,~787~nm and a Ti:Sph tunable laser for the wavelength of $\lambda$=~801.5 nm and 806.5 nm. The power density in any case was $\rm{I=~0.6~ mW/mm^2}$. By using a
$\lambda/4$ plate and a liquid crystal half-wave variable retarder in tandem, the polarization of the laser can be switched between left- and right-circular polarization. The laser  incidence is  normal to the surface of the sample and focused within a spot size of 1.0 mm diameter. The PL is directed to a linear polarizer prisms to select the left- and right-circular polarizations, by rotating the prism. The PL signal is collected and detected by using an imaging spectrometer with a resolution of $0.025~{\rm nm}$ equipped with a thermoelectric-cooled camera as detector.

\medskip
\textbf{Acknowledgements} \par
We would like to thank E. Ontiveros, F. Ram\'\i rez-Jacobo, L. E. Guevara-Mac\'\i as and J. Gonzalez-Fortuna for their skillful technical support. This work was supported by Consejo Nacional de Ciencia y Tecnolog\'\i a (Grant No. FC 2016-02-2093).



\begin{table}
	\caption{Parameters of the DQWs structures used in the present work.}
	\centering
	\begin{tabular}[htbp]{@{}lll@{}}
		\hline
		Sample & AlGaAs barrier doping level & QWs thicknesses $d_1$, $d_2$\\
		\hline
		Symmetric & Si doped n= $6\times 10^{18}\rm{cm^{-3}}$  & 11.9 nm, 11.9 nm\\
		Asymmetric& Si doped n= $6\times 10^{18}\rm{cm^{-3}}$ & 23.7 nm, 11.9 nm \\
		Asymmetric& Be doped p= $5\times 10^{19}\rm{cm^{-3}}$  & 23.7 nm, 11.9 nm \\
		\hline
	\end{tabular}
  \label{Table_1}
\end{table}

\end{document}